**Using Curriculum Theory to Inform Approaches to Generative AI in Schools**

Myke Healy, BA (Hons), B.Ed, MEd, OCT [he/him]
Doctoral Program, Werklund School of Education, University of Calgary
August 6, 2023

**Abstract**

In an educational landscape dramatically altered by the swift proliferation of Large Language Models (LLMs), this essay interrogates the urgent pedagogical modifications required in secondary schooling. Anchored in Madeline Grumet's triadic framework of curriculum inquiry, the study delineates the multifaceted relationships between Generative AI and Elliot Eisner's concepts of explicit, implicit, and null curricula. It scrutinizes the logistical and ethical challenges—such as the reliability of AI detectors—that educators confront when attempting to assimilate this nascent technology into long-standing curricular structures. Engaging with Ted Aoki's theory of the "zone of between," the essay illuminates the dilemmas educators face in reconciling prescriptive curricular aims with the fluid realities of classroom life, all within the context of an educational milieu in constant flux due to Generative AI. The paper culminates in a reflective analysis by the researcher, identifying avenues for further scholarly investigation within each of Grumet's constitutive strands of curriculum theory, thereby providing a roadmap for future research on Generative AI's transformative impact on educational practice.

**Keywords**

artificial intelligence, generative AI, curriculum theory, Elliot Eisner, Ted Aoki,  Madeline Grumet, curriculum inquiry, LLM, large language models, ChatGPT, academic dishonesty, AI-giarism, explicit implicit null curriculum

**Using Curriculum Theory to Inform Approaches to Generative AI in Schools**

Education has never faced a disruptive technology with such a rapid exponential adoption curve as we have with Generative AI. Prior technological advances that impacted education—social media, smartphones, learning management systems, the internet, laptops, photocopiers, calculators, TV, radio, and blackboards—all changed pedagogy. Yet none of these technologies unleashed their transformative power with the same sudden force or required such rapid adaptation by schools. To illustrate the broader cultural uptake, ChatGPT was released on November 30, 2022, climbed to one million users in five days, 100 million users by February 2023, and is now regularly fielding 1.6 billion visits a month (Brown, 2023). Much has been written within education, with much hand-wringing, about how Generative AI (GenAI) will unleash a tide of plagiarism, or 'AI-giarism,' as dubbed by Cecilia Chan (2023). While a landscape of widespread academic dishonesty is assured in the coming academic year, that is not the focus of this inquiry. My primary interests lie in understanding how pedagogy in secondary schools must adapt to GenAI, how this profound change will happen, and the fraught path to get there. This paper will explore how curriculum theory informs the transformative impact of Generative AI on the relationships between curriculum, teachers, and students. As a framing reference for the investigation, I will explore Madeline Grumet's (2008) three strands of curriculum inquiry, using the contention that the release of ChatGPT is a significant inflection point in the "study of the curriculum object as an event" (p. 138). To illustrate Grumet's remaining two strands, the scholarship of Elliot Eisner and Ted Aoki will highlight the connection between GenAI and curriculum theory, followed by my perspective as a researcher. I will close with a brief exploration of where further study is needed in each of Grumet's three strands to assist future scholarship in this area.

## The Study of the Curriculum Phenomenon as a Cultural Object

The second of Grumet's (2008) strands of curriculum inquiry relates to complex curricular phenomena "with a social history, anchored in ideology, and nested in layers of meaning that call for clarification and interpretation" (pp. 137–138). Thinking of how Generative AI will disrupt traditional ways of teaching is a compelling thought experiment. To my mind, Grumet's call for clarifying and interpreting the "cultural object" (p. 137) of GenAI can best be done through the lens of Eisner's (2002) explicit, implicit, and null curriculums.

### The Explicit Curriculum

The explicit curriculum related to GenAI has yet to be written for secondary schools in Canada, and knowing the glacial pace of provincial curriculum change, it may not happen for several years. Indeed, should the brightest Canadian scholars sit down to write curriculum on this topic, how could it be done for such a rapidly changing target? The challenge of capturing technological innovation in an explicit curriculum is understandable. Innovative educators will take it upon themselves to teach about GenAI in the 2023–2024 academic year, notwithstanding a dearth of provincial curriculum. Such instruction should, at minimum, happen in computer studies courses, for even if GenAI is not currently in provincial curricula, it is captured within broader objectives. For example, a curriculum expectation in the Ontario Ministry of Education Grade 11 Computer Studies course (last updated in 2008) notes: "By the end of this course, students will demonstrate an understanding of emerging areas of research in computer science" (Computer Studies, 2008, p. 45). I appreciate Eisner's (2004) argument about how curriculum objectives like this both help and hinder with their clarity:

Educational objectives clearly and specifically stated can hamper as well as help the ends of instruction, and that an unexamined belief in curriculum…can easily become dogma which in fact may hinder the very functions the concept was originally designed to serve.

(p. 85)

Teachers who take the "dogma" approach and avoid the topic of Generative AI this coming year do students a disservice.

In the year ahead, Generative AI instruction will dwell in Ted Aoki's (2012) "zone of

between," that liminal space between broad provincial curriculum-as-plan expectations and the

teacher's curriculum-as-lived experience (p. 161). The coming lived experience for educators

will be impacted by a steady stream of news updates related to the ever-widening providence of

GenAI functionality and the challenging questions about how education and society must adapt.

The connected teacher who feels duty-bound to address the cultural object of GenAI will need to create their own explicit curriculum on the fly while simultaneously navigating the lived experience of a connected human impacted by the very technology they are attempting to teach.

Aoki (2012) captures this feeling as "living in tensionality—a tensionality that emerges, in part,

from indwelling in a zone between two curriculum worlds" (p. 161). This tension will be a challenge and may become part of the implicit curriculum as students watch teachers and administrators struggle to navigate the changing landscape of AI in schools.

**The Implicit Curriculum**

How schools approach and use Generative AI in the coming year is a compelling example of the implicit curriculum at work. Eisner (2002) notes that "the implicit curriculum of the school is what it teaches because of the kind of place it is. And the school is that kind of place through the ancillary consequences of various approaches to teaching" (p. 97). One can imagine the dichotomy between the implicit curriculum of a school in 2023 that actively discusses the use, ethics, and opportunities of Generative AI compared to a school that tries to ban, block, and punish its use. New York City schools banned ChatGPT in January 2023 and then lifted the ban four months later, demonstrating the challenges of creating and maintaining policies related to disruptive technology (Sapienza, 2023).

The nascent field of AI plagiarism detection software is a minefield of implicit curriculum messaging. When an online detector falsely determined that an entire class had used ChatGPT to write their final assignments in May 2023, a Texas A&M University professor threatened to give the entire class an incomplete grade, impacting graduation. One student caught up in the false-positive dragnet remarked about a feeling of betrayal, noting, "The thought of my hard work not being acknowledged and my character being questioned...It just really frustrated me." (Verma, 2023). Subsequent journalist investigations have found that the leading AI-detector, GPTZero, scored the US Constitution and the Bible as "most likely generated by AI," among other absurdities in this emerging field (Edwards, 2023). A false positive leading to academic dishonesty charges against an innocent student sends damaging implicit curriculum messaging around school-sponsored injustice and elevates algorithms over human trust. In contrast, administrators could infuse implicit justice messaging into their academic integrity approach. As an example, educational leaders could apply the classic principle of justice: it is more acceptable for 100 students to commit plagiarism via ChatGPT than for one student to suffer a wrongful accusation due to AI-detection software.

Eisner (2002) asserts that a school's implicit curriculum "can teach a host of intellectual and social virtues" (p. 95). I see this as an opportunity, for even if teachers stumble through the Aokian "zone of between" while navigating their curriculum-based lessons and what they understand about GenAI, doing so honestly and openly implicitly teaches students a way of being with change. Seeing teachers admit what Aoki (2012) terms "meaningful striving and struggling" (p. 164) in relation to new AI tools demonstrates a commitment to understanding new paradigms as a human responsibility. Accepting and teaching about Generative AI reveals respect for our students, as these tools will arguably shape the course of their lives more than our own. My optimistic viewpoint is that the coming tumult in education will allow teachers to role model Eisnerian "social virtues," in this case, a commitment to understanding the new and adapting to change, and respect for preparing students for a changing future. Should everything go horribly wrong, the implicit curriculum might also teach that algorithms can replace humans, all conventional forms of assessment are useless, and school is woefully insufficient to teach students what they need to know for a future AI-infused world.

**The Null Curriculum**

What schools choose to neglect in the coming years of disruptive technology innovation will be informative. Eisner (2002) advocates that scholars interrogate what is absent from school curricula "in order to reassure ourselves that these omissions were not a result of ignorance but a product of choice" (p. 98). There is an understandable amount of ignorance surrounding Generative AI in schools, and educators must take on the personal task of learning the technology, assisted by administrators who can provide timely training. Some schools will choose to ignore GenAI, wishing—ostrich-like—that student take-home writing can remain a reliable proxy for student thinking and understanding, a paradigm that has existed in differing forms since antiquity. In my view, the act of relegating Generative AI to the null curriculum is tantamount to willful blindness. It risks fostering unchecked academic dishonesty and, in extreme cases, undermining the credibility of traditional credentialing. Indeed, a recent experiment by Maya Bodnick (2023) found that the large language model GPT-4 earned "a respectable 3.57 GPA" when used to complete freshman-year assignments at Harvard. She starkly concludes that "AI isn't just coming for the college essay; it's coming for the cerebral class" (Bodnick, 2023). We are entering a time of educational arbitrage, where "AI tools are increasingly being used in productive, disruptive, imaginative, and disquieting ways before regulators, K–12 schools, and universities can determine how best to implement guardrails and guidelines" (Healy, 2023a). One of the best guardrails is to move teaching with and about GenAI from the 'null' to the 'explicit' side of the curriculum ledger.

In a second scenario, let us assume institutions appreciate the gravity of the technology and GenAI finds its place in the explicit and implicit curriculum in schools. Some educational leaders' actions may align with Franklin Bobbitt's 1918 characterization that an "Education that prepares for life is one that prepares definitely and adequately for these specific activities" (Bobbitt, 2004, p. 11). Prioritizing GenAI in preparation for the future may inadvertently push certain disciplines to the periphery, or worse, into the null curriculum, given the limited number of school hours available. In such an outcome, disciplines with performative learning and assessments, such as the arts, may gain new stature as uniquely human endeavours worthy of time and attention in schools. Taken even further, a world in which much of our present school activities can be conducted via Generative AI may move us closer to John Dewey's principle of the school as "primarily a social institution" where education "is a process of living and not a preparation for future living" (Dewey, 2004, p. 19). In such a utopian scenario, the distinctly human-only skills of empathy, compassion, service, and citizenship that often dwell in the null curriculum may find renewed prominence in the explicit and implicit curriculums.

### The Study of Curriculum in the Perspective of the Researcher

True to the autobiographical nature of curriculum scholarship, Madeline Grumet notes that "Curriculum inquiry requires a recapitulation of the researcher's own history of experience and associations with the object to be studied" (Grumet et al., 2008, p. 138). On this count, Gruemet's characterization of being "saturated and shaped" (p. 138) by Generative AI is an apt personal description for me. I started experimenting with AI image generators in September 2022 and moved to ChatGPT soon after its release on November 30, 2023. Since these early weeks, I have read voraciously on the topic and found writers on Substack and Medium particularly helpful, most notably the work of Ethan Mollick (2023) and Alberto Romero (2023). In terms of social media, TikTok is proving the best way to follow how *students* are coming to understand and use GenAI tools. A growing body of scholarly work is emerging on Generative AI, and I find Semantic Scholar an effective AI tool that sends alerts whenever new studies are published.

As a school administrator, I gave my first of multiple presentations to our school's faculty on January 26, 2023, followed by a one-hour presentation to the School's board of governors on February 23, 2023. For many, these presentations were the first time they learned and saw what Generative AI tools could do, and the feedback was a mix of excitement, concern, fear, and wonderment. My early presentations focused on academic integrity and the inherent bias in AI image generators that routinely created images rife with stereotypes. Throughout 2023, engineers continued to add parameters to the large language models to address many of these concerns. ChatGPT quickly became a staple of conversation at the staff lunch table, and we facilitated faculty PD sessions on Generative AI on May 4 and June 20, 2023. At the close of the 2022–2023 academic year, I encouraged teachers to use ChatGPT over the summer and learn about the technology. I also struck and chaired the first three meetings of our AI Task Force, where we discussed three preliminary questions:

1. How will our teaching, assessment, and evaluation practices adapt in a learning environment with widely accessible generative AI tools?

2. How must our academic integrity policies change to address generative AI?

3. Should we use generative AI tools to assist in writing report card comments? (Healy, 2023b)

Out of these meetings, we developed a framework for using ChatGPT for report card writing and drafted academic integrity guidelines related to generative AI use. Beyond our institution, I presented at the Conference of Independent Schools Ontario meeting on May 16, 2023, and facilitated two in-depth national workshops for colleagues at the Canadian Accredited Independent Schools Leadership Institute on July 5 and 6, 2023.

The requirement to present material helps me synthesize key insights into representative quotes, images, and slide decks. Knowing the Generative AI landscape will look substantively different in one, five, and ten years, I have archived and time-stamped my July 2023 presentations. In my future dissertation, I will reflect on this early material and understanding of the GenAI curriculum circa 2023.

## Areas for Future Study

I propose three areas of future study related to Generative AI and curriculum theory, following Grumet's (2008) three strands of curriculum inquiry.

### Future research related to the study of the curriculum object as an event

Conducting historical research into contemporary scholarship at the outset of other major disruptive educational technologies may yield helpful insights. Some initial commentary suggests that the reaction to Generative AI is merely another manifestation of Sisyphean technology panic. Many strongly contest this viewpoint, convinced that GenAI will bring about societal disruption on a scale comparable to the internet and do so at a far quicker pace. Looking into scholarship from 1983–1994 about the future impact of the internet will be informative, both from a cultural and research methodology perspective, as will determining what prescient questions were asked or not asked.

**Future research related to the study of the Curriculum Phenomenon as a Cultural Object**

There is no shortage of research and source material related to Generative AI as a current cultural object. Indeed, a more apt metaphor is a daily firehose of content on this topic that requires triage, sifting, and categorizing for future use. In order to track and build a corpus of curriculum on the topic, I am curating a list of annotated resources, which in future years will also serve as a historical record of how educators' understanding of Generative AI changed over time.

**Future research related to the study of Curriculum in the Perspective of the Researcher:**

One concern I have is that when Canadian schools have adapted to reality with Generative AI, this scholarship will not be relevant. I can appreciate that articulating a current phenomenological understanding in the moment can serve as a source for future reflection. By articulating my engagement in this field since the fall of 2022, I am recording these early months to inform my candidacy and dissertation work in 2026–2027.

## Conclusion

Studying an emerging technology tool poses a number of challenges, not the least of which is having your thinking continually challenged by new use cases, scholarly work, and cultural commentary. In two or three years, this essay may feel 'quaint' in its understanding of how generative artificial intelligence will impact schools. By recording early understandings, archiving introductory presentations, creating a repository of early scholarship, and connecting initial insights to established curriculum theory, my goal is to establish a phenomenological foundation for future study in the field.